\documentclass{elsart}
\usepackage{mathrsfs}
\usepackage{bbm}
\usepackage{amsmath,amssymb}
\usepackage{theorem}
\usepackage{url}
\usepackage{version}
\usepackage{alltt}

\def\N {\ensuremath{\mathbb{N}}}

\def\F {\ensuremath{\mathbb{F}}}

\def\K {\ensuremath{\mathbb{K}}}

\def\M {\ensuremath{\mathsf{M}}}

\def\E{\textsf{Expand}}
\def\D{\textsf{Decomp}}

\def\mul{\mathrm{mul}}
\def\rev{\mathrm{rev}}

\newtheorem{Theorem}{Theorem}

\def\proof{\textsc{Proof.} }
\def\foorp{\hfill$\square$}

\begin{document}

\begin{frontmatter}
\title{Fast Conversion Algorithms \\ for Orthogonal Polynomials}
\author{Alin Bostan} 
\ead{Alin.Bostan@inria.fr} \qquad
\author{Bruno Salvy} 
\ead{Bruno.Salvy@inria.fr} 
\address{Algorithms Project, INRIA Rocquencourt \\ 78153 Le Chesnay Cedex France} \medskip
and
\author{{\'E}ric Schost} 
\ead{eschost@uwo.ca}
\address{ORCCA and Computer Science Department, Middlesex College, \\
University of Western Ontario, London, Canada}

\begin{abstract}
We discuss efficient conversion algorithms for orthogonal polynomials. We describe a known 
conversion algorithm from an arbitrary orthogonal basis to the monomial basis,
and deduce a new algorithm of the same complexity for the converse operation.
\end{abstract}


\maketitle 
\end{frontmatter}


\section{Introduction}

Let $(a_i)_{i \ge 1}$, $(b_i)_{i \ge 1}$ and $(c_i)_{i \ge 1}$ be
sequences with entries in a field~$\K$. We can then define the
sequence $(F_i)_{i \geq 0}$ of orthogonal polynomials in $\K[x]$ by
$F_{-1} = 0$, $F_0 = 1$ and for $i \ge 1$ by the second order
recurrence
\begin{equation}\label{eq:1}
F_{i} = (a_{i} x+b_{i}) F_{i-1} + c_{i} F_{i-2}.
\end{equation}
Following standard conventions, we require that $a_i c_i$ is non-zero
for all $i \geq 1$; in particular, $F_i$ has degree $i$ for all $i\geq
0$ and $(F_i)_{i \geq 0}$ forms a basis of the $\K$-vector space
$\K[x]$.

Basic algorithmic questions are then to perform efficiently the base
changes between the basis $(F_i)_{i \geq 0}$ and the monomial basis
$(x^i)_{i \geq 0}$. More precisely, for $n \in \mathbb{N}\setminus
\{0\}$, we study the following problems.

\medskip
\begin{description}
\item[Expansion Problem ($\E_n$).] Given $\alpha_0,\dots,\alpha_{n-1}
\in \K$, compute the coefficients on the monomial basis of the
polynomial $A$ defined by the map
\begin{equation}\label{eq:exp}
	[\alpha_0,\dots,\alpha_{n-1}] \mapsto A=\sum_{i=0}^{n-1} \alpha_i
F_i\end{equation} 
\item[Decomposition Problem ($\D_n$).] Conversely, given the
coefficients of $A$ on the monomial basis, recover the coefficients
$\alpha_0,\dots,\alpha_{n-1}$ in the decomposition~\eqref{eq:exp} of
$A$ as a linear combination of the $F_i$'s.
\end{description}

\medskip
For $i,j \ge 0$, let $F_{i,j}$ be the coefficient of $x^i$ in $F_j$,
and let ${\bf F}_n$ be the $n\times n$ matrix with entries
$[F_{i,j}]_{0 \le i,j < n}$. Problem $\E_n$ amounts to multiplying the
matrix ${\bf F}_n$ by the vector $[\alpha_0,\dots,\alpha_{n-1}]^t$;
hence, the inverse map $\D_n$ is well-defined, since ${\bf F}_n$ is an
upper-triangular matrix whose $i$-th diagonal entry $F_{i,i} = a_1 a_2
\cdots a_i$ is non-zero. As we will see, the {\it dual problem}
(multiplying the matrix $\mathbf{F}_n^t$ by a vector), denoted
$\E_n^t$, plays an important role as well.

Naive algorithms work in complexity $O(n^2)$ for both problems $\E_n$
and $\D_n$. Faster algorithms are already known, see details below on
prior work. The only new result in this article is the second part of
Theorem~\ref{Theo:1} below; it concerns fast computation of the map $\D_n$. 

As usual, we denote by $\M$ a \emph{multiplication time} function,
such that polynomials of degree less than $n$ in $\K[x]$ can be
multiplied in $\M(n)$ operations in~$\K$, when written in the monomial
basis. Besides, we impose the usual super-linearity conditions
of~\cite[Chap.~8]{GaGe99}.  Using Fast Fourier Transform algorithms,
$\M(n)$ can be taken in $O(n \log(n))$ over fields with suitable roots
of unity, and in $O(n \log(n)\log\log(n))$ over any
field~\cite{ScSt71,CaKa91}.

\begin{Theorem}\label{Theo:1}
  Problems \emph{$\E_n$} and \emph{$\D_n$} can be solved in
  $O(\M(n)\log(n))$ arithmetic operations in $\K$.
\end{Theorem}

The asymptotic estimates of Theorem~\ref{Theo:1} also hold for
conversions between any arbitrary orthogonal bases, using the monomial
basis in an intermediate step. In conjunction with FFT algorithms for
polynomial multiplication, Theorem~\ref{Theo:1} shows that all such
base changes can be performed in \emph{nearly linear time.}


\paragraph*{Previous work.} 
Fast algorithms are known for problems closely related to Problem
$\E_n$. {}From these, one could readily infer fast algorithms for
Problem $\E_n$ itself.

In~\cite{PoStTa98}, the question is the computation of the values
$$[\alpha_0,\dots,\alpha_{n-1}] \mapsto \big [~ \sum_{i=0}^{n-1}
\alpha_i F_i(x_j) ~\big ]_{0 \le j < n},$$ where the $x_j$ are the
Chebyshev points $x_j=\cos(j\pi/(n-1))$. This is done by expanding
$\sum_{i=0}^{n-1} \alpha_i F_i$ on the Chebyshev basis and applying a
discrete cosine transform. The article~\cite{DrHeRo97} studies the
transposed problem: computing the map
\begin{equation}\label{eq:DrHeRo97}[\alpha_0,\dots,\alpha_{n-1}] \mapsto \big [~ \sum_{i=0}^{n-1} \alpha_i F_j(x_i) ~\big
]_{0 \le j < n}.\end{equation} The algorithm in~\cite{DrHeRo97} is
(roughly, see~\cite{PoStTa98} for details) the transpose of the one
in~\cite{PoStTa98}: it applies a transposed multipoint evaluation, then 
a transposed conversion, to either the monomial or the Chebyshev basis.

Regarding problem $\D_n$ to the best of our knowledge, no $O(\M(n)
\log(n))$ algorithm has appeared before, except for particular
families of polynomials, like Legendre~\cite{Frumkin95},
Chebyshev~\cite{Pan98} and Hermite~\cite{LeRoCh07}. In the case of
arbitrary orthogonal polynomials, the best complexity result we are
aware of is due to Heinig~\cite{Heinig01}, who gives a
$O(\M(n)\log^2(n))$ algorithm for solving inhomogeneous linear systems
with matrix ${\bf F}_n^t {\bf F}_n$. {}From this, it is possible to
deduce an algorithm of the same cost for Problem $\D_n$.

In~\cite{PoStTa98}, one sees mentions of left and right inverses for
the related problem
$$[\alpha_0,\dots,\alpha_{n-1}] \mapsto \big [~ \sum_{i=0}^{n-1}
\alpha_i F_i(x_j) ~\big ]_{0 \le j < 2n-1}.$$ In~\cite{LeRoCh07}, the
inverse of the map~\eqref{eq:DrHeRo97} is discussed: when $(x_i)$ are
the roots of~$F_{n}$, Gauss' quadrature formula shows that this map is
orthogonal, so that inversion reduces to transposition. In other
cases, approximate solutions are given.

The various algorithms mentioned up to now have costs
$O(\M(n)\log(n))$ or $O(\M(n)\log^2(n))$.  In~\cite{BoSaSc08b}, we
give algorithms of lower cost $O(\M(n))$ for many classical orthogonal
polynomials (Jacobi, Hermite, Laguerre, \dots), for both Problems
$\E_n$ and $\D_n$.


\paragraph*{Main ideas.}
Here is a brief description of the strategy used to obtain the
complexity estimate of Theorem~\ref{Theo:1}. The complete treatment
with detailed algorithms is given in Sections~\ref{sec:expand}
and~\ref{sec:decomp}. Three main ingredients are used: $(i)$ a
$O(\M(n)\log(n))$ algorithm for Problem $\E_n$; $(ii)$ the
transposition principle; $(iii)$ the Favard-Shohat theorem.

We first recast \eqref{eq:1} into the matrix recurrence
$[F_i,F_{i+1}]^t = {\bf M}^{(i)}(x) [F_{i-1},F_{i}]^t$, where ${\bf
M}^{(i)}$ is a $2\times 2$ polynomial matrix. Problem $\E_n$ then
amounts to computing $T=\alpha_0{\bf M}^{(0)} + \alpha_1{\bf M}^{(1)}
{\bf M}^{(0)} + \cdots + \alpha_{n-1} {\bf M}^{(n-1)} \cdots {\bf
M}^{(0)}$.  This is done by using a divide-and-conquer algorithm
similar to the one in~\cite[Th.~2.4]{Gerhard00} for the conversions
between Newton and monomial bases. Assuming for simplicity that $n$ is
even, we rely on the decomposition $T= T_0 + T_1 \, {\bf
M}^{(\frac{n}{2}-1)} \cdots {\bf M}^{(0)}$, with
\begin{eqnarray*}
T_0 &=& \alpha_0{\bf M}^{(0)} + \cdots + \alpha_{\frac{n}{2}
1} {\bf M}^{( \frac{n}{2} -1)} \cdots {\bf
M}^{(0)}\\[1mm] T_1 &=& \alpha_{ \frac{n}{2}}{\bf
M}^{( \frac{n}{2})} + \cdots + \alpha_{n-1} {\bf
M}^{(n-1)} \cdots {\bf M}^{( \frac{n}{2}
)}.\end{eqnarray*} In Section~\ref{sec:expand}, a slightly
different but more efficient version of this algorithm is given.

An algorithmic theorem called the {\it transposition
principle}~\cite[Th.~13.20]{BuClSh97} states that the existence of an
algorithm of cost $O(\M(n)\log(n))$ for $\E_n$ implies the existence
of another one \emph{with the same cost} for the dual problem
$\E^t_n$. We use an effective version of the principle, allowing to
design the transposed algorithm in a straightforward manner starting
from the direct one.

Then, the Favard-Shohat theorem~\cite{Favard35,Shohat36} ensures the
existence of an inner product $\langle \, , \, \rangle$ on the space
$\K[x]$ with respect to which the sequence $(F_n)_n$ is an orthogonal
basis. This implies the matrix equality $ {\bf F}_n^t \, {\bf H}_n\,
{\bf F}_n \,=\, {\bf D}_n$, where ${\bf H}_n = [h_{i,j}]_{0 \le i,j <
n}$ is the Gram matrix with $h_{i,j} = \langle x^i, x^j \rangle$ and $
{\bf D}_n$ is an invertible diagonal matrix. Its equivalent form ${\bf
F}_n^{-1} \, = \, {\bf D}_n^{-1}\, {\bf F}_n^t \, {\bf H}_n$ shows
that, once $ {\bf H}_n$ and $ {\bf D}_n$ are determined, Problem
$\D_n$ amounts to the computation of the map $w\in \K^n \mapsto {\bf
F}_n^t w \in \K^n$, that is, to solving $\E^t_n$. Finally, a
constructive version of the Favard-Shohat theorem shows that
determining the Gram matrix ${\bf H}_n$ can be reduced to two
instances of Problem $\E_n$.

In summary, by the Favard-Shohat theorem, $\D_n$ is reduced to $\E_n$
and $\E_n^t$, which can be solved in $O(\M(n)\log(n))$, by a direct
divide-and-conquer algorithm for the first and the transposition
principle for the second.


\section{Expansion Problem}\label{sec:expand}
We first describe the conversion from the orthogonal basis to the
monomial one, and its transpose. The content of this section is mostly
already known. However, our algorithm for the inverse operation rests
crucially on these conversions, so we prefer to make them explicit.

In the following, we always suppose for simplicity that the number of
unknown coefficients $n$ is a power of two. For a polynomial $F$ of degree less
than $m$, $m \geq 1$, we denote by $\rev(F,m)=x^{m-1} F(1/x)$ the reversal
of $F$.

{\bf Expansion from an orthogonal basis.}
Given $\alpha_0,\dots,\alpha_{n-1}$, we compute here the expansion
on the monomial basis of
$$A=\alpha_0 F_0 + \cdots + \alpha_{n-1} F_{n-1}.$$ The ideas 
are classical; our presentation is taken from~\cite{PoStTa98}. However,
our use of ``classical'' fast multiplication techniques avoids the
need of precomputed constants arising in~\cite{PoStTa98}, and holds
over any field. For $i \ge 0$, define the transition matrix
$${\bf M}^{(i,i+1)} = \left [ \begin{matrix} 0 & 1 \\ c_{i+1} & a_{i+1} x + b_{i+1} \end{matrix} \right ],$$
so that we have
$$\left [ \begin{matrix} F_{i} \\ F_{i+1} \end{matrix} \right ]
 = {\bf M}^{(i,i+1)} \left [ \begin{matrix} F_{i-1} \\ F_{i} \end{matrix} \right ].$$
For  $j > i$, let ${\bf M}^{(i,j)}= {\bf M}^{(j-1,j)}  {\bf M}^{(j-2,j-1)} \cdots {\bf M}^{(i,i+1)};$
for $i=j$, ${\bf M}^{(i,j)}$ is the $2\times 2$ identity matrix.
It follows that we have 
$$\left [ \begin{matrix} F_{j-1} \\ F_{j} \end{matrix} \right ]
 = {\bf M}^{(i,j)} \left [ \begin{matrix} F_{i-1} \\ F_{i} \end{matrix} \right ];$$
besides, for $\ell \ge j \ge i$, we have the associativity relation
${\bf M}^{(i,\ell)} = {\bf M}^{(j,\ell)} {\bf M}^{(i,j)}.$
We can then rewrite $A$ as
$$\left [ \begin{matrix} A \end{matrix} \right ]
= \left [ \begin{matrix} \alpha_0 & \alpha_1 \end{matrix} \right ]
            \left [ \begin{matrix} F_0 \\ F_1 \end{matrix} \right ]  + \cdots + 
            \left [ \begin{matrix} \alpha_{n-2} & \alpha_{n-1} \end{matrix} \right ]
            \left [ \begin{matrix} F_{n-2} \\ F_{n-1} \end{matrix} \right ],$$
where the sum has $n/2$ terms. We deduce the equalities
$$
\left [ \begin{matrix} A \end{matrix} \right ] =  
\left [ \begin{matrix} \alpha_0 & \alpha_1 \end{matrix} \right ] {\bf M}^{(1,1)}
            \left [ \begin{matrix} F_0 \\ F_1 \end{matrix} \right ]  + \cdots + 
            \left [ \begin{matrix} \alpha_{n-2} & \alpha_{n-1} \end{matrix} \right ] {\bf M}^{(1,n-1)}
            \left [ \begin{matrix} F_0 \\ F_1 \end{matrix} \right ] \\
= {\bf B} \left [ \begin{matrix} F_0 \\ F_1 \end{matrix} \right ],$$
where $\bf B$ is the $1 \times 2$ matrix
$\displaystyle{
{\bf B} = \sum_{i=0}^{n/2-1} \left [ \begin{matrix} \alpha_{2i} \
    \alpha_{2i+1}  \end{matrix} \right ] {\bf M}^{(1,2i+1)}.
}$

The computation of $A$ is thus reduced to that of the matrix ${\bf B}.$
Write $n'=n/2$. Following~\cite{Stone73} and~\cite{KoSt73}, we build
the {\it subproduct tree} associated to the transition matrices ${\bf
M}^{(j,i)}$. This is a complete binary tree having
$d=\log_2(n)=\log_2(n')+1$ rows of nodes labeled as follows:
\begin{itemize}
\item the leaves of the tree are labeled by the matrices
  ${\bf L}^{(d-1,i)}={\bf M}^{(2i+1,2i+3)}$, for $i=0,\dots,n'-1$;

\medskip

\item for $j=0,\dots,d-2$, there are $2^j$ nodes of depth $j$ and
  the $(1+i)$-th one is indexed by the matrix ${\bf L}^{(j,i)} =
  {\bf L}^{(j+1,2i+1)}{\bf L}^{(j+1,2i)}$, for $0\leq i \leq 2^j -1$.
\end{itemize}
The entries $L^{(j,i)}_{u,v}$ of ${\bf L}^{(j,i)}$ have degrees at
most $2^{d-j}-2+u+v$, with $0 \le u,v \le 1$. An easy induction also
shows that for $j=0,\dots,d-1$ and $i=0,\dots,2^j-1$, we have the
equality
$${\bf L}^{(j,i)} = {\bf M}^{(2^{d-j}i+1,\, 2^{d-j}(i+1)+1)}.$$ The
cost of computing all matrices in the tree is $O(\M(n)\log(n))$, as
in~\cite[Chapter 10]{GaGe99}. Then, to compute ${\bf B}$, we go up the
subproduct tree and perform linear combinations along the way: we
maintain a family of $1\times 2$ vectors ${\bf
v}^{(j,i)}=[v_0^{(j,i)}~v_1^{(j,i)}]$, with $j=0,\dots,d-1$ and
$i=0,\dots,2^j-1$, such that 
\begin{equation}\label{eq:linearcomb}
 {\bf
v}^{(d-1,i)}=[\alpha_{2i}~\alpha_{2i+1}] \quad\text{and}\quad {\bf
v}^{(j,i)} = {\bf v}^{(j+1,2i)} + {\bf v}^{(j+1,2i+1)} {\bf
L}^{(j+1,2i)}.
\end{equation}
The overall cost is again $O(\M(n)\log(n))$. 

Remark that not all the nodes of the complete subproduct tree are
actually needed in this algorithm. Indeed, its rightmost branch
containing ${\bf L}^{(j,2^j-1)}$ for $0\leq j \leq d-1$ is not
necessary in the computation described in
Equation~\eqref{eq:linearcomb}.

\begin{figure}[t]
\begin{center} \fbox{
\begin{minipage}{9 cm}
\begin{center} $\underline{\E({\bf a}, {\bf b}, {\bf c}, A,n)}$ \end{center}
\vspace{0.5cm}

\textbf{Input:} $A = \sum_{i=0}^{n-1} \alpha_i x^i$
and ${\bf a}, {\bf b}, {\bf c}$\\
\textbf{Output:} $\sum_{i=0}^{n-1} \alpha_i F_i$\\
 
\begin{tabbing}
\quad \= \quad \= \quad \= \quad \kill
${\bf L}^{(j,i)} \leftarrow \textsf{SubproductTree}({\bf a}, {\bf b}, {\bf c}, n)$\\
{\bf for} $i=0,\dots,2^{d-1}-1$ {\bf do} \\
\> $v_0^{(d-1,i)} \leftarrow  \alpha_{2i}$\\
\> $v_1^{(d-1,i)} \leftarrow  \alpha_{2i+1}$\\
{\bf for} $j=d-2,\dots,0$ {\bf do}\\
\> {\bf for} $i=0,\dots,2^j-1$ {\bf do}\\
\> \> $v^{(j,i)}_0 \leftarrow v^{(j+1,2i)}_{0}\ +\ v^{(j+1,2i+1)}_0\ L^{(j+1,2i)}_{0,0}\ +\ v^{(j+1,2i+1)}_1\ L^{(j+1,2i)}_{1,0}$\\
\> \> $v^{(j,i)}_1 \leftarrow v^{(j+1,2i)}_{1}\ +\ v^{(j+1,2i+1)}_0\ L^{(j+1,2i)}_{0,1}\ +\ v^{(j+1,2i+1)}_1\ L^{(j+1,2i)}_{1,1}$\\
{\bf return} $v^{(0,0)}_0 F_0 + v^{(0,0)}_1 F_1$
\end{tabbing}
\end{minipage}
}\end{center}
\caption{Algorithm solving Problem $\E_n$}
\label{algo:expand}
\end{figure}

In the pseudo-code in Figure~\ref{algo:expand}, we make all scalar
operations explicit, so as to make the transposition process easier in
the next paragraph. Starting from the sequences ${\bf
a}=(a_1,\ldots,a_{n-1}), {\bf b}=(b_1,\ldots,b_{n-1}), {\bf
c}=(c_1,\ldots,c_{n-1})$, the subroutine \textsf{SubproductTree}(${\bf
a, b, c}, n$) computes the matrices ${\bf M}^{(i,i+1)}$ for $0\leq i
\leq n-2$, then the matrices ${\bf L}^{(j,i)}$ for $1\leq j \leq d-1$
and $0\leq i \leq 2^j -2$.

\paragraph*{Transposed expansion.} 
Let $r,s \geq 1$ and let ${\bf M}$ be a $r \times s$ matrix with
entries in $\K$. The {\it transposition
principle}~\cite[Th.~13.20]{BuClSh97} states that the existence of an
algorithm for the matrix-vector product $b \mapsto {\bf M} b$ implies
the existence of an algorithm with the same cost, up to $O(r+s)$
operations, to perform the transposed matrix-vector product $c \mapsto
{\bf M}^t c$. This paragraph gives the transposed version of the
conversion algorithm above: a similar algorithm is given
in~\cite{DrHeRo97}, but our derivation is substantially more compact.

A fundamental operation is transposed polynomial multiplication. For
$k$ in $\N$, let $\K[x]_k$ be the $\K$-vector space of polynomials of
degree less than $k$. Then, for $B$ in $\K[x]$ of degree $m$, we let
$\mul(.,B,k)$ be the multiplication-by-$B$ operator, defined over
$\K[x]_k$; its image lies in $\K[x]_{k+m}$.

\begin{figure}[t]
\begin{center} \fbox{
\begin{minipage}{9 cm}
\begin{center} $\underline{\E^t({\bf a}, {\bf b}, {\bf c}, A,n)}$ \end{center}
\vspace{0.5cm}
\textbf{Input:} $A=\sum_{i=0}^{n-1} \alpha_i x^i$ and ${\bf a}, {\bf b}, {\bf c}$\\
\textbf{Output:} $(\,\sum_{i=0}^{n-1} \alpha_i {\rm coeff}(F_j,i)\,)_{j=0,\dots,n-1} $\\
 
\begin{tabbing}
\quad \= \quad \= \quad\quad\quad\quad\quad \= \quad \= \quad \kill
${\bf L}^{(j,i)} \leftarrow \textsf{SubproductTree}({\bf a}, {\bf b}, {\bf c},n)$\\
$v^{(0,0)}_0 \leftarrow {\rm mul}^t(A, F_0, \delta_{d,0})$\\
$v^{(0,0)}_1 \leftarrow {\rm mul}^t(A, F_1, \delta'_{d,0})$\\[1mm]
{\bf for} $j=0,\dots,d-2$ {\bf do}\\
\> {\bf for} $i=2^j-1,\dots,0$ {\bf do}\\
\> \> $v^{(j+1,2i)}_{0}$ \> $\leftarrow$ \> $v^{(j,i)}_0 \bmod x^{\delta_{d,j+1}}$\\[1mm]
\> \> $v^{(j+1,2i)}_{1}$ \> $\leftarrow$ \> $v^{(j,i)}_1 \bmod x^{\delta'_{d,j+1}}$\\[1mm]
\> \> $v^{(j+1,2i+1)}_{0}$ \> $\leftarrow$ \> ${\rm mul}^t(v^{(j,i)}_0, L^{(j+1,2i)}_{0,0}, \delta_{d,j+1})$\\[1mm]
\> \>  \> $+$ \> ${\rm mul}^t(v^{(j,i)}_1, L^{(j+1,2i)}_{0,1}, \delta_{d,j+1})$\\[1mm]
\> \> $v^{(j+1,2i+1)}_{1}$ \> $\leftarrow$ \> ${\rm mul}^t(v^{(j,i)}_0, L^{(j+1,2i)}_{1,0}, \delta'_{d,j+1})$\\[1mm]
\> \>  \> $+$ \> ${\rm mul}^t(v^{(j,i)}_1, L^{(j+1,2i)}_{1,1}, \delta'_{d,j+1})$\\[1mm]
{\bf return} $v^{(d-1,0)}_0,v^{(d-1,0)}_1,\dots,v^{(d-1,2^{d-1}-1)}_0,v^{(d-1,2^{d-1}-1)}_1$
\end{tabbing}
\end{minipage}
}\end{center}
\caption{Algorithm solving Problem $\E_n^t$}
\label{algo:texpand}
\end{figure}

The transpose of this map is denoted by $\mul^t(.,B,k)$; by
identifying $\K[x]_k$ with its dual, one sees that $\mul^t(.,B,k)$
maps $\K[x]_{k+m}$ to $\K[x]_{k}$. In~\cite{BoLeSc03,HaQuZi04},
details of the transposed versions of plain, Karatsuba and FFT
multiplications are given, with a cost matching that of the direct
product. Without using such techniques, writing down the
multiplication matrix shows that $\mul^t(.,B,k)$ is $$A \in
\K[x]_{k+m} \mapsto (A\ \rev(B,m+1) \bmod x^{k+m}){\rm~div~}x^{m} \in
\K[x]_k.$$ Using standard multiplication algorithms, this formulation
leads to slower algorithms than those
of~\cite{BoLeSc03,HaQuZi04}. However, here $k$ and $m$ are of the same
order of magnitude, and only a constant factor is lost.

Using this tool, the transposed expansion algorithm in
Figure~\ref{algo:texpand} is obtained by reversing the flow of the
direct one in Figure~\ref{algo:expand}. The loops are traversed in
opposite order. Then, the operation ${\bf v}^{(j,i)} = {\bf
v}^{(j+1,2i)} + {\bf v}^{(j+1,2i+1)} {\bf L}^{(j+1,2i)}$ in the inner
loop is replaced by a truncated copy of ${\bf v}^{(j,i)}$ into ${\bf
v}^{(j+1,2i)}$ and a transposed matrix-vector product, where
polynomial multiplications are replaced by transposed
multiplications. To perform truncations and transposed
multiplications, we need information on the degrees of the polynomials
involved. By induction, we get the following inequalities, for $j=0,\dots,d-1$
and $i=0,\dots,2^j-1$, $$\deg(v^{(j,i)}_0) <
\delta_{d,j}=\max(0,2^{d-j}-3)+1, \quad \deg(v^{(j,i)}_1) <
\delta'_{d,j}=2^{d-j}-1.$$ This information enables us to write the
transposed algorithm in Figure~\ref{algo:texpand}. Using either the
transposition principle or a direct analysis, one sees that the cost
of this algorithm is $O(\M(n)\log(n))$.


\section{Decomposition Problem}\label{sec:decomp}

The Favard-Shohat theorem~\cite{Favard35,Shohat36}, see
also~\cite[Theorem 4.4]{Chihara78}, asserts that for $(F_i)$ as
in~\eqref{eq:1}, there exists a linear form $L: \K[x] \to\K$ for which
$(F_i)$ is {\it formally orthogonal}, in the sense that, for $i \ge
1$,
\begin{equation*}
L(F_i F_j) =0 \quad\text{for}\quad 0 \le j < i, \quad L(F_i^2)\ne 0.
\end{equation*} 
The linear form $L$ is specified by its {\it moments} $L(x^i)$, for $i
\ge 0$, or equivalently by the generating series $$S_L = \sum_{i \ge
0} L(x^i) x^i \in \K[[x]].$$

For completeness, we give in the following theorem a self-contained,
constructive, proof of this classical result, showing how to compute
truncations of $S_L$. The proof is inspired by the presentation
in~\cite[Section~3]{Flajolet80}.

\begin{Theorem}
Let $(F_i)$ be the sequence satisfying $F_{-1} = 0, F_0=1$ and recurrence~\eqref{eq:1}.
 Define the sequence $(G_i)$ by  $G_{-1} = 0$, $G_0=1$ and, for $i \ge 1$
  \begin{equation*} \label{eq:2}
   G_{i} = (a_{i+1} x+b_{i+1}) G_{i-1} + c_{i+1} G_{i-2}.
  \end{equation*}
Then, there exists a $\K$-linear form $L: \K[x] \to\K$ such that
	\begin{equation}\label{eq:L}
		L(F_i F_j) =0 \quad\text{for}\quad i \neq j , \quad \text{and} \quad L(F_i^2) = (-1)^i \, \frac{c_2\cdots c_{i+1}}{a_{i+1}} \quad\text{for}\quad i \ge 0.
	\end{equation}
Moreover, for any $i \geq 1$, the following equality holds between truncated series in
$\K[[x]]$:
	\begin{equation} \label{eq:momentSeries}
	\frac{{\rm rev}(G_{i-1},i)}{{\rm rev}(F_i,i+1)} =   \sum_{i \ge 0} L(x^i) x^i  \mod x^{2i}.
	\end{equation}
\end{Theorem}

\proof For $i\ge 0$, write $F_i^\star={{\rm rev}(F_i,i+1)}$ and
$G_i^\star={{\rm rev}(G_i,i+1)}$. Let also define $F_{-1}^\star = G_{-1}^\star = 0$. These polynomials satisfy the recurrences
$$F_{i}^\star = (a_{i}+b_{i}x) F^\star_{i-1} + c_{i}x^2 F^\star_{i-2},  \quad G_{i}^\star = 
(a_{i+1} +b_{i+1}x) G^\star_{i-1} + c_{i+1}x^2 G^\star_{i-2},$$
for $i \ge 1$, which 
can be recast into the matrix form
$$\left [ \begin{matrix}  F^\star_{i} & G^\star_{i-1} \\F^\star_{i+1} & 
G^\star_{i}\end{matrix}\right ]
= \left [ \begin{matrix}  0 & 1 \\ c_{i+1}x^2 & (a_{i+1}+b_{i+1}x)  \end{matrix}\right ]
  \left [ \begin{matrix}  F^\star_{i-1} & G^\star_{i-2} \\F^\star_{i} & 
G^\star_{i-1}\end{matrix}\right ].$$
Taking determinants, we deduce that for $i \ge 1$ the following identity holds
$$
\frac{G^\star_{i}}{F^\star_{i+1}}-\frac{G^\star_{i-1}}{F^\star_{i}}
= -c_{i+1} \frac{F^\star_{i-1}}{F^\star_{i+1}} x^2 \left (
\frac{G^\star_{i-1}}{F^\star_{i}}-\frac{G^\star_{i-2}}{F^\star_{i-1}}
\right ).
$$
Applying it to $i,i-1,\ldots$ and denoting $\gamma_i = c_2 \cdots c_i$, we get  that for $i\ge 1$,
\begin{equation} \label{eq:takeDet}
 \frac{G^\star_{i}}{F^\star_{i+1}} -  \frac{G^\star_{i-1}}{F^\star_{i}} = (-1)^i 
\frac{\gamma_{i+1}}{F^\star_{i} F^\star_{i+1}} x^{2i}.
\end{equation}
A separate check shows that Equation~\eqref{eq:takeDet} also holds for $i=0$. 

For $i\ge 0$, $F_i^\star$ has constant coefficient
$\delta_i=a_1\cdots a_i$, which is non-zero, and is thus invertible in
$\K[[x]]$. Since the $\gamma_{i+1}$ are non-zero as well,
Equation~\eqref{eq:takeDet} shows that the sequence
$G^\star_{i}/F^\star_{i+1}$ is Cauchy and thus convergent in
$\K[[x]]$. Besides, if we let $S$ be its limit, summing up
Equation~\eqref{eq:takeDet} for $i,i+1,\ldots$ yields
\begin{equation} \label{eq:S}
S = \frac{G^\star_{i-1}}{F^\star_{i}} + (-1)^i
  \frac{\gamma_{i+1}}{\delta_i \delta_{i+1}} x^{2i} \mod x^{2i+1}, \quad \text{for} \quad i\geq 0.
\end{equation}
 Write $S=\sum_{i \ge 0} \ell_i x^i$ and define the linear form $L$ on
  $\K[x]$ by $L(x^i)=\ell_i$. Then Equation~\eqref{eq:momentSeries} is
  a direct consequence of~\eqref{eq:S}.
  
  For $i \ge 0$, equating coefficients of $x^i, \ldots, x^{2i-1}$ and
$x^{2i}$ in Equation~\eqref{eq:S} multiplied by ${F^\star_{i}}$
implies $L(F_i x^j) = 0$ for $i < j$ and $L(F_i x^i) = (-1)^i
\gamma_{i+1}/\delta_{i+1}$.  By linearity, this shows that $L$
also satisfies Equality~\eqref{eq:L}.  \foorp

\begin{figure}[t]
\begin{center} \fbox{
\begin{minipage}{9 cm}
\begin{center} $\underline{\D({\bf a}, {\bf b}, {\bf c}, A, n)}$ \end{center}
\vspace{0.5cm}

\textbf{Input:} 
$A = \sum_{i=0}^{n-1} u_i x^i $ and ${\bf a}, {\bf b}, {\bf c}$\\
\textbf{Output:} $\alpha_0,\ldots,\alpha_{n-1}$ such that $A= \sum_{i=0}^{n-1} \alpha_i F_i$ \\
 
\begin{tabbing}
${\bf a}' \leftarrow \textsf{cat}({\bf a}, 1)$ \\ [1mm]
${\bf b}' \leftarrow \textsf{cat}({\bf b}, 1)$ \\ [1mm]
${\bf c}' \leftarrow \textsf{cat}({\bf c}, 1)$\\ [1mm]
$F\leftarrow \E({\bf a}', {\bf b}, {\bf c}, x^{n},n+1)$ \\ [1mm]
$G\leftarrow \E({\bf Sa}', {\bf Sb}', {\bf Sc}', x^{n-1}, n)$ \\ [1mm]
$Q \leftarrow \rev(G,n)/\rev(F,n+1) \mod x^{2n-1}$\\ [1mm]
$V\leftarrow \mul^t(Q,A,n)$ \\  [1mm]
$w  \leftarrow  \E_n^t ({\bf a},{\bf b}, {\bf c}, V,n)$ \\  [1mm]
$d_i  \leftarrow  (-1)^i  c_2 \cdots c_{i+1} / a_{i+1}$, for $0\leq i <n$\\ [1mm]
$\alpha_i \leftarrow  w_i / d_i$ for  $0\leq i <n$, where $w=[w_0,\ldots,w_{n-1}]^t$\\[1mm]
{\bf return} $\alpha_0,\ldots,\alpha_{n-1}$
\end{tabbing}
\end{minipage}
}\end{center}
\caption{Algorithm solving Problem $\D_n$}
\label{algo:decomp}
\end{figure}

\medskip
\paragraph*{Proof of Theorem~\ref{Theo:1}.}
We can now prove the second part of Theorem~\ref{Theo:1}, dealing with expansions in the monomial basis. 
The corresponding algorithm is given in Figure~\ref{algo:decomp}.

We first compute $(L(x^i))_{i < 2n-1}$. To do this, we start from the
sequences ${\bf a}, {\bf b}$ and ${\bf c}$ to which we add the element
$1$, in order to make the polynomial $F = F_{n}$ well-defined (any
non-zero choice would do). We then use the algorithm \textsf{Expand}
of the previous section to compute $G = G_{n-1}$ and $F = F_{n}$ and
we determine the power series expansion $\rev(G_{n-1},n)/\rev(F_n,n+1)
\bmod x^{2n-1}$.  The first step takes $O(\M(n)\log(n))$ operations,
and the second one $O(\M(n))$ using Newton iteration~\cite[Chap.~9]{GaGe99}.
In the pseudo-code we use the notation ${\bf Sx}$ for the shifted
sequence $(x_{i+1})$ of ${\bf x} = (x_{i})$ and the notation
\textsf{cat} for concatenation.

Consider finally the matrix ${\bf F}_n$ defined in the introduction,
and let ${\bf H}_n=[H_{i,j}]_{0 \le i,j < n}$ be the $n\times n$
Hankel matrix with $H_{i,j} = L(x^{i+j})$. Let next ${\bf D}_n$ be the
diagonal matrix of size $n$, with $D_i = L(F_i^2)$. We deduce the
factorization $$ {\bf F}_n^t \, {\bf H}_n\, {\bf F}_n \,=\, {\bf D}_n,
\quad\text{or}\quad {\bf F}_n^{-1} \, = \, {\bf D}_n^{-1}\, {\bf
F}_n^t \, {\bf H}_n.$$
Equation~\eqref{eq:L} shows that one can compute the entries of
${\bf D}_n$ in $O(n)$ operations. 

At this stage, all elements of ${\bf D}_n$ and ${\bf H}_n$ are
known. Right-multiplication of ${\bf H}_n$ by the coefficient vector
of a polynomial $A \in \K[x]_n$ amounts to the transposed
multiplication of the polynomial $Q = \sum_{i=0}^{2n-2} L(x^i) x^i$ by
$A$, that can be performed in time $\M(n)+O(n)$. Using the transposed expansion algorithm
$\textsf{Expand}^t$ of the previous section, multiplication by ${\bf
F}_n^t$ costs $O(\M(n)\log(n))$. Finally, multiplying by ${\bf
D}_n^{-1}$ takes linear time. This concludes the proof of Theorem~\ref{Theo:1}.

\smallskip\noindent{\bf Acknowledgments.} This work was supported in part by the French National Agency for Research (ANR Project ``Gecko'') and the Microsoft Research-INRIA Joint Centre.

\bibliographystyle{plain}
\bibliography{BoSaSc08b}

\end{document}